# Fluorescence-detected Wavepacket Interferometry reveals time-varying Exciton Relaxation Pathways in single Light-Harvesting Complexes

Stephan Wiesneth[1], Paul Recknagel[1], Alastair T. Gardiner[3]
Richard Cogdell[2], Richard Hildner[4], Jürgen Köhler[1,5,6]

*[1]Spectroscopy of soft Matter, University of Bayreuth, 95440 Bayreuth, Germany*
*[2]School of Molecular Biosciences, Glasgow University, Glasgow G12 8QQ, UK.*
*[3] Institute of Microbiology of the Czech Academy of Sciences, 379 01 Třeboň, Czech Republic*
*[4]Zernike Institute for Advanced Materials, University of Groningen, Nijenborgh 3, 9747 AG Groningen, The Netherlands.*
*[5]Bavarian Polymer Institute, University of Bayreuth, 95440 Bayreuth, Germany*
*[6]Bayreuther Institut für Makromolekülforschung (BIMF), 95440 Bayreuth, Germany*

corresponding author:

Jürgen Köhler, Spectroscopy of soft Matter, University of Bayreuth,

95440 Bayreuth, Germany,

phone: +49 921 554000,

email: juergen.koehler@uni-bayreuth.de

## Abstract

Photosynthesis relies on efficient energy relaxation within the excited-state manifold of pigment-protein complexes. Since the protein scaffold is rather flexible, the resulting energetic and structural disorder gives rise to a complex excited-state energy level structure that fluctuates on all time scales. Although the impact of such fluctuations on relaxation processes is known, the precise exciton states involved in relaxation as well as the nature of the vibrational modes driving relaxation are under debate. Here single pigment-protein complexes from a photosynthetic purple bacterium are excited with two identical ultrashort phase-locked pulses producing two exciton wave packets that can interfere. This leads to a modulation of the emission intensity as a function of the delay time between the pulses that fades out within about $\approx$100 fs due to fluctuating environments on those time scales. For several single complexes we find variations of the interference patterns on time scale of several 10 s that reveal fluctuations in the energy relaxation pathways towards the lowest-energy exciton states. This relaxation is driven by temporal variations in the coupling between electronic excitations and low-frequency vibrational modes.

## Introduction

In photosynthesis photons are absorbed by light-harvesting complexes (LHCs) and the energy of the excited states is transferred rapidly to the photochemical reaction centre (RC) where an electron transfer chain is initiated[1–3]. The LHCs accommodate chromophores that are properly positioned with respect to each other by a protein scaffold. The close proximity of the chromophores leads to strong electronic coupling and the formation of delocalized excited states (excitons), which correspond to coherent superpositions of the excited states of the individual chromophores. However, the protein matrix can be considered as a dielectric environment (bath) that randomly fluctuates on essentially all time scales, leading to unsynchronized modulations of both the excited state energies of the individual chromophores (site energies) and the coupling strengths between the chromophores, usually referred to as dynamical diagonal and off-diagonal disorder, respectively. Although coherences are destroyed by such fluctuations, the progress in non-linear ultrafast time-resolved spectroscopy, and in particular the development of two-dimensional electronic spectroscopy (2DES), has allowed long-lived quantum coherences to be observed and boosted a debate about the character of these coherences (electronic, vibrational, or vibronic) and their possible relevance for the function of the LHCs[4–14]. Obviously, without perturbations the system would oscillate without any net energy transfer, whereas too much disorder might trap an electronic excitation and hamper exciton mobility towards the RC. Hence, efficient energy transfer requires both coherence and decoherence, and it seems that nature established a balance, between these two, to regulate and direct energy transport in photosynthetic light harvesting [12,15–17].

However, 2DES requires high excitation intensities to be able to retrieve the (higher order) non-linear signals. Hence, often substantial excited-state absorption is present in the signal and/or impulsive Raman transitions are induced that create vibrational wave packets in the electronic ground state. Such contributions are difficult to separate from the signals related to (electronic/vibronic) coherence in the lowest-energy exciton manifold that is relevant for function in light harvesting. Moreover, due to the structural heterogeneity of the proteins and the unsynchronized character of the conformational fluctuations, details of ultrafast dynamics are washed out by ensemble averaging. For instance, fluctuations of relaxation pathways on the time scale of seconds are expected, since conformational reorganization of the protein scaffold occurs on that time scale[18–24], but those cannot be resolved with bulk techniques. It is therefore difficult to retrieve details about precise relaxation pathways with 2DES, although complex polarization-controlled schemes allirowed some insights to be gained[25–27].

By combining ultrafast spectroscopy and single-molecule techniques this problem can be overcome. To the best of our knowledge, the first ultrafast spectroscopy on single LHCs was reported by the van Hulst group[28]. The experiments were conducted on the peripheral light-harvesting complexes (LH2) from purple bacteria. Details of the atomistic structures of these complexes depend on the bacterial species and on their growth conditions[2,29–37] yet a common feature is the organisation of the BChl *a* molecules, one ring of weakly interacting BChl *a* molecule and one ring of closely interacting BChl *a* molecules, giving rise to two absorption bands at 800 nm (referred to as B800) and 850 nm (B850), respectively, see also Fig.S1 in section 1 of the supporting information. The time-resolved experiments reported[28] were conducted on single LH2 complexes from the species *Rps. acidophila*. Using a fluorescence-detected two-colour double-pump approach an oscillatory signal was revealed that was attributed to electronic quantum coherences between the B800 and the B850 states. Variations in the oscillatory signal on slow (seconds) time scales were ascribed to changes in relaxation path-ways between B800 and B850 as well as within the B850 states due to structural fluctuations within the binding pocket of the chromophores. Later the fluctuations of the relaxation times on similar slow timescales were resolved on single LH2 complexes by pump-probe spectroscopy[38]. However, in both studies the precise pathways were not accessible, i.e., there was no information on the B850 exciton states that were involved in the (time-varying) relaxation.

Here, we resolve this relaxation of excitons and variations of relaxation pathways on timescales of seconds within the B850 exciton band of single LH2 complexes from *Rps. acidophila.* As suggested by Fleming and coworkers already more than a decade ago[39], single LH2 complexes are excited with two identical, phase-locked ultrafast pulses of very low intensity avoiding singlet-singlet annihilation and other non-linear effects. The excitation scheme produces two excitonic wave packets that can interfere giving rise to a modulation of the emission signal as a function of the delay (and phase) between the pulses. These modulations fade out within roughly 100 fs. Systematically changing the lock frequency of the pulses allowed us to associate temporal variations of the interference patterns with fluctuations of the exciton relaxation pathways involving specific B850 states.

## Experimental

Sample preparation

LH2 complexes from *Rhodopseudomonas (Rps.) acidophila* (strain 10050) were isolated and purified as detailed in[40]. Stock solutions of LH2 complexes in buffer (20 mM Tris/HCl, pH 8.0, 0.1 % LDAO) were stored at -80 °C. For control experiments on ensembles of LH2 complexes the stock solution was diluted to a concentration of about 1 nM, and for studying individual complexes it was further diluted to a concentration of about 10 pM. For both preparations 2 % (w/w) polyvinyl alcohol (PVA) was added to the buffer solution in the final dilution step. A drop of 20 µL of the final solution is spin-coated on a coverslip, and the sample was immediately placed in a home-built vacuum chamber to diminish bleaching.

Optical Experiment

The optical experiments were carried out on a home-built setup. A titanium-sapphire laser-system (Griffin-10-WT, KMLabs) was used to generate ultrashort broadband excitation pulses. For exciting the B800 and B850 bands of LH2, the pulse spectrum was adjusted to a center-wavelength of 820 nm (12195 $cm^{-1}$) and a bandwidth of 80 nm (1178 $cm^{-1}$), which results in a transform-limited pulse width of about 20 fs (full width at half maximum, FWHM) after compression. For the experiment, where only the B850 band was excited, the spectral components below 820 nm were blocked by a razor edge in the pulse shaper. This resulted in a pulse centred at 840 nm (11905 $cm^{-1}$) with a bandwidth of 40 nm (560 $cm^{-1}$), corresponding to a transform-limited pulse width of about 40 fs (FWHM) after compression.

A pulse shaper (MIIPS-HD, BiophotonicSolutions) was used for pulse compression and generation of time-delayed, phase-locked pulse-pairs by simultaneous amplitude and phase shaping, implementing the amplitude modulation mask $A(\omega) = |cos(0.5[(\omega -$ $\Omega_L)\tau + \Delta\varphi])|$ and the phase modulation mask $\phi(\omega) = 0.5\pi sgn[cos(0.5[(\omega - \Omega_L)\tau + \Delta\varphi])]$ in a spatial light modulator (SLM) [41]. An example for the transmitted spectral profile of the excitation light after passing the spatial light modulator (SLM) is shown in Fig.S2 in section 2 of the supporting information. The additional application of a phase mask yields two pulses separated by a time $\tau$ in the time domain. Variation of the transmission characteristics of the SLM in combination with appropriate phase masks allowed the delay between the pulses to be scanned in increments of 6 fs. The excitation light was focused by a water-immersion objective (NA=1.2) into the sample plane, and the fluorescence was collected by the same objective. The signal passed a long pass filter (edge at 885 nm, AHF) to suppress residual laser light and was directed towards a single-photon counting module (SPCM-AQR-16, Perkin Elmer). To obtain a reference signal the emission upon excitation with a single transform-limited pulse was recorded between subsequent pulse pairs. Note that pulse compression was performed using multi-photon intra-pulse interference phase scans[42] with a non-linear BBO crystal in the focal plane of the water-immersion objective, such that transform-limited pulses were obtained in the sample plane.

The advantage of preparing the pulses with an SLM instead of a delay line is that it gives access to control the mutual phases for distinct spectral components in the two pulses, and that it gives full control of the excitation intensities. Here, the time-averaged excitation intensity was set to about $200\frac{W}{cm^2}$ for all delays. This corresponds to a fluence of $1.0 \times 10^{13}\frac{photons}{pulse \cdot cm^2}$, or equivalently to 1 excitation per 10 pulses (using $\sigma_{B800} = 1.0 \times 10^{-14} cm^2$ for the B800 absorption cross section[43]), which puts us clearly into the linear excitation regime of the complex. All experiments were carried out at room temperature with the sample held under vacuum.

## Results

LH2 complexes from *Rps. acidophila* were excited by two ultrashort transform-limited pulses that were phase-locked at a frequency $\Omega_L$, and that were separated in time by a delay $\tau$. The response from LH2 complexes was studied for three different experimental conditions referred to hereafter as scenarios A, B, and C. In Fig.1A-C the spectral profiles of the pulses are shown by the grey shaded areas together with the ensemble absorption spectrum of LH2, which features the typical broad absorption bands at about 800 nm (yellow) and 850 nm (red). For scenario A the spectral profile of the pulses overlaps with the full B800 band and a large part of the B850 band, and the lock frequency $\Omega_L$ corresponded to 12,500 $cm^{-1}$ (800 nm). For scenario B, $\Omega_L$ was kept at 12,500 $cm^{-1}$ but the spectral profile of the pulses was cut-off to exclude the B800 excitations. Finally, for scenario C, as before the B800 excitations were excluded but $\Omega_L$ was set to 11,900 $cm^{-1}$ (840 nm).

The response of LH2 from *Rps. acidophila* for the three scenarios is compared in Fig.1 as a function of the delay between the pulses. Each column of the figure represents from left to right an example for one of the scenarios A, B, and C, as also illustrated at the top of the figure. The responses from a macroscopic ensemble of LH2 complexes for the three scenarios are shown in Fig.1D-F for phase shifts between pulses of $\Delta\varphi = 0$ (red lines) as well as for $\Delta\varphi = \pi$ (blue lines) as a control experiment. The detected fluorescence intensities show clear modulations with relative depths of 42% (Fig.2D), 58% (Fig.2E), and 17% (Fig.2F) that fade out within 200 fs. The inverted oscillatory response upon excitation with delayed and $\pi$-phase-shifted pulses reveals the interference character of the modulations.

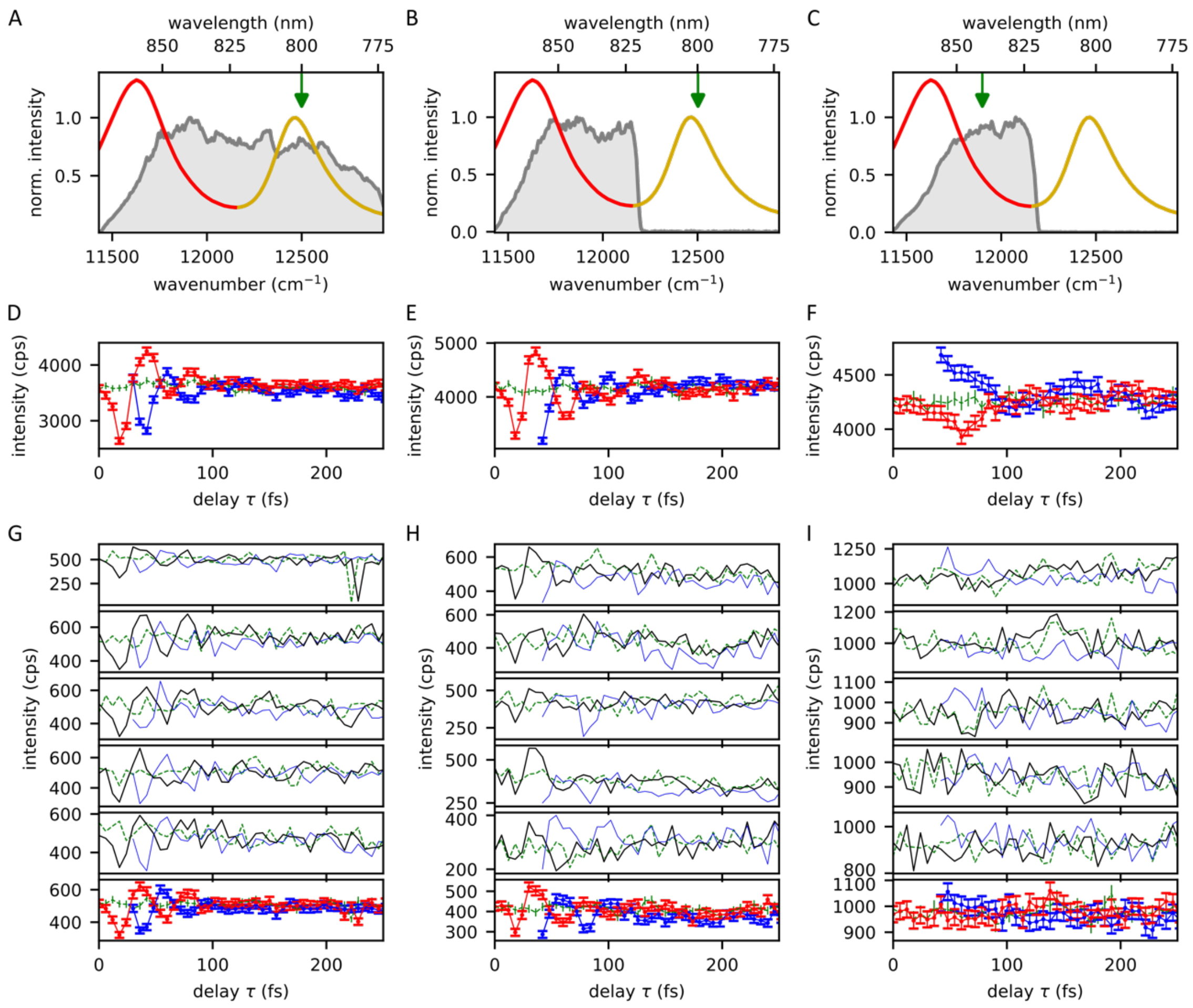

Fig.1: A) - C) The spectral profiles of the laser pulses in the frequency domain (grey shaded areas) overlaid with an ensemble absorption spectrum of LH2 from *Rps. acidophila*, where the yellow (red) part refers to the B800 (B850) absorption band. The lock frequencies for the experiments in the respective columns (e.g. A, D, G) are indicated by the green arrows. D) - F) Response from a macroscopic ensemble of LH2 complexes upon excitation with two phase-locked pulses as a function of the pulse delay $\tau$ for the excitation conditions shown in parts A) - C). G) - I) Examples for the response from single LH2 complexes upon excitation with two phase-locked pulses as a function of the pulse delay $\tau$ for the excitation conditions shown in parts A) - C). The black curves are individual scans with a phase difference of $\Delta\varphi = 0$ between the pulses and the blue curves correspond to scans with a phase difference of $\Delta\varphi = \pi$ between the two pulses. The red (blue) traces at the bottom correspond to the average of the black (blue) traces shown in the same panel, and the green lines correspond to the reference signal obtained from a single transform-limited pulse. The error bars take the shot noise and fluctuations of the laser intensity into account. For all experiments the excitation intensity was $200 \frac{W}{cm^2}$ during data acquisition. The signal is given as intensity in counts per second (cps). The acquisition time for a single scan was 60 s.

The responses from representative examples of single LH2 complexes for the same excitation conditions are shown in Fig.1G-I (see Fig.S3 for a single-complex bleaching trace). Each panel features five scans of the delay ($\Delta\varphi = 0$ black lines; $\Delta\varphi = \pi$ blue lines), the resulting average at the bottom ($\Delta\varphi = 0$ red lines; $\Delta\varphi = \pi$ blue lines), and the reference signal (green lines). Both the individual scans as well as the averages feature a clear modulation of the detected emission signal that fades out on a time scale of 50 - 100 fs. For an unbiased analysis of the modulation traces the data from individual complexes served as input for a Fourier transform, which yields for each individual delay trace a single modulation frequency $f$ (an example is provided in Fig.S4). For scenario A, 25 individual complexes were studied, and the modulation frequencies covered a range from 14 THz - 25 THz (corresponding to 470 $cm^{-1}$ - 830 $cm^{-1}$) centred at a mean value of 20.3 THz (675 $cm^{-1}$), Fig.2A. For scenario B, 15 individual complexes were studied, and the modulation frequencies were distributed between 10 THz and 27 THz (corresponding to 330 $cm^{-1}$ - 900 $cm^{-1}$) centred at a mean value of 21.4 THz (710 $cm^{-1}$), Fig.2B. Finally, for scenario C, 25 individual complexes were studied and the modulation frequencies covered the range from 6.5 THz - 13.1 THz (corresponding to 220 $cm^{-1}$ – 440 $cm^{-1}$) centred at a mean value of 8.7 THz (290 $cm^{-1}$), Fig.2C. Interestingly, while each trace from an individual complex yielded a single frequency, we found for several complexes in each of the scenarios that this frequency changed between subsequent traces of the same complex by up to 8 THz (twice the resolution of 4 THz determined by the inverse of the maximum pulse delay of 250 fs) on a time scale of some 10 seconds.

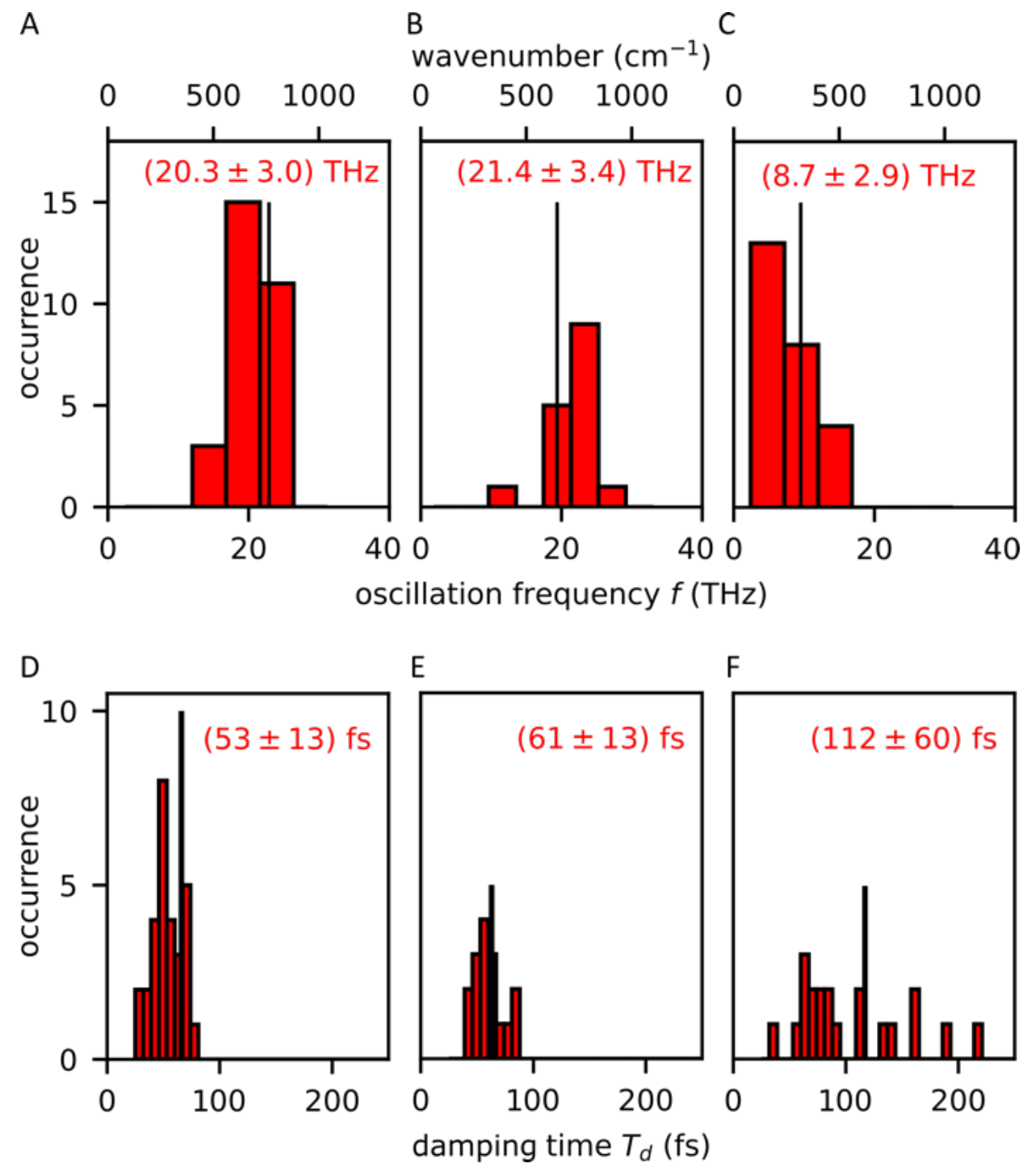

Fig.2: A) - C) Distributions of the modulation frequencies $f$ of the emission signal from individual LH2 complexes for the three experimental scenarios A, B, and C, respectively. D) - F) Distributions of the damping times $T_d$ from individual LH2 complexes, for D) and E) also shown on an expanded scale in the insets. The statistical parameters (mean ± standard deviation) of the distributions are given in the figures. The vertical black lines indicate the ensemble values that amount to A) 22.9 THz, B) 19.4 THz, C) 9.5 THz, D), 66 fs, E) 63 fs, F) 117 fs, respectively.

In order to obtain a measure for the damping times $T_d$ the traces from the individual complexes were fitted by a single, exponentially damped cosine function using $I(\tau) = C_1 * cos(2\pi f \cdot \tau + \varphi)e^{\frac{-\tau}{T_d}} + C_2$, where $\varphi$ denotes a phase factor, and $C_1$ and $C_2$ are constants that refer to the amplitude and the background, respectively. The modulation frequencies $f$ were fixed to the values obtained before by the Fourier transforms. The damping times $T_d$ were distributed as shown in Fig.2D-F with characteristic parameters (mean ± sdev) of (53 ± 13) fs for scenario A, (61 ± 13) fs for scenario B, and for scenario C (112 ± 60) fs.

## Discussion

The electronically excited states of the B850 molecules in the LH2 complexes are usually described by the Frenkel exciton model[44–46], which yields as exciton states $|k\rangle = \sum_{n=1}^{N} c_{nk}|n\rangle$ , i.e., coherent superposition states of electronic excitations localized on molecule *n* and weighted by coefficients $c_{nk}$. In the simplest approximation (i.e. identical site energies for all B850 BChl *a* molecules, and without energetic disorder) this yields the well-known ladder of exciton states numbered *k = 0*, *k = ± 1*, ... *k = ± 8, k = 9* [47–51]. Due to the head-to-tail arrangement of the transition-dipole moments of the B850 BChl *a* molecules in the plane of a ring nearly all of the oscillator strength is concentrated in the *k = ± 1* exciton states causing the main absorption band at about 850 nm[48,50].

For the three scenarios reported above the first pulse creates a coherent superposition between the ground state and the optically allowed excited states that are resonant with the laser spectrum, i.e., an exciton wave packet. For scenario A, the spectral bandwidth of the pulses covers the B800 band and a large part of the B850 band, whereas the excitation of the B800 states is excluded for the scenarios B and C. For the discussion of the response of a single LH2 complex upon excitation with a phase-locked ultrashort pulse pair, we have to consider all states $|\Psi_i\rangle$ in the B800 and B850 bands that possess oscillator strength with the electronic ground state (see Fig. S5 for an illustration based on three essential states, a B800 state ($|\Psi_{800}\rangle$), a higher exciton state in the B850 band ($|\Psi_{850}^{h}\rangle$), and an emitting state in the B850 band ($|\Psi_{850}^{e}\rangle$)). It should be noted that all states are to some extent a mixture of (mainly localized) B800 and (mainly delocalized) B850 states and include also some charge-transfer (CT) character[52,53]. The wave packet created by excitation with the laser pulses can then be written $|\Psi_{WP}(t)\rangle = \sum_i c_i(t)|\Psi_i\rangle$ with time-dependent coefficients $c_i(t)$. Upon excitation the initial composition of the wave packet is determined by the transition-dipole moments of the excited states involved, and the electric field amplitude at the specific absorption wavelengths, in particular $c_{i=B800}(t=0)=0$ for scenarios B and C. After the first pulse the wave packet evolves under field-free conditions and each electronic state $|\Psi_i\rangle$ accumulates a time-dependent phase $\phi(t)=-i(\Omega_L-\omega_i)t$ , where $\Omega_L$ denotes the lock frequency, and $\hbar\omega_i$ refers to the energy of the state $|\Psi_i\rangle$[54]. Given the non-stationary character of the wave packet, the energy oscillates between the excited states, and in parallel the population relaxes toward the lower lying exciton states, see also Fig.S5 in section 5 of the supporting information. The relaxation might cause some loss of phase memory and might add additional phase shifts $e^{i\zeta_i}$ to the states $|\Psi_i\rangle$[28]. However, from 2DES measurements it was concluded that interaction with inter- and intra-molecular vibrations, resonant with the energy separation between these states, preserves coherence between the $|\Psi_i\rangle$ states for several 100 fs[27,55]. After a delay time $\tau$ the second pulse creates another exciton wave packet (B800/B850 wave packet for scenario A, for scenarios B and C only a B850 wave packet), with phases $\phi(t+\tau)=-i\big((\Omega_L-\omega_i)(t+\tau)+\Delta\varphi\big)$. Here, the phase $\Delta\varphi$ refers to a fixed phase difference between the two pulses. Except for the control experiments (blue lines in Fig.1), where $\Delta\varphi=\pi$ was used, for all other experiments the phase difference between the two pulses was $\Delta\varphi=0$. Hence, each Fourier component (frequency) of the superposition state created by the second pulse has a well-defined phase difference with the same Fourier component (frequency) generated after the first pulse. This phase difference depends on both the energy difference $\Omega_L-\omega_i$ between the lock frequency and the energy of the state $|\Psi_i\rangle$, and on the delay time $\tau$. The two wave packets can interfere and the total probability amplitude for finding the system in the emitting states after interaction with the pulses and relaxation can be enhanced or reduced due to constructive or destructive interference. Since the spontaneous emission is directly related to the population of the emitting states, the fluorescence intensity depends on the delay $\tau$, the phase difference $\Delta\varphi$ between the pulses, and the accumulated phase of the wavepacket during the time interval between pulses.

However, random environmental fluctuations modulate the site energies of the chromophores and the energies of the states $|\Psi_i\rangle$ should better be expressed as $\hbar\omega_i = \hbar\left(\omega_0 + \overline{\delta\omega_i} + \delta\omega_i(t)\right)$, where $\omega_0$ corresponds to the average site energy, $\overline{\delta\omega_\iota}$ to a static offset and $\delta\omega_i(t)$ to a time-dependent random fluctuation with mean zero. As a consequence, the site energies of the chromophores during the first pulse and during the second pulse are not identical, and the resulting phases are then $\phi(t) = -i\left(\Omega_L - \left(\omega_0 + \overline{\delta\omega_i} + \zeta + \delta\omega_i(t)\right)\right)t$ and $\phi(t+\tau) = -i\left(\left(\Omega_L - \left(\omega_0 + \overline{\delta\omega_i} + \delta\omega_i(t+\tau)\right)\right)(t+\tau) + \varphi\right)$. Hence, interferences can be observed as long as the sum of the autocorrelations of the fluctuations $\sum_i\langle\delta\omega_i(0)\cdot\delta\omega_i(\tau)\rangle_t \neq 0$ does not vanish. Here $\langle\ldots\rangle_t$ denotes time averaging. This implies that the individual realizations of $\delta\omega_i(t)$ must remain sufficiently similar during the delay time $\tau$ between the pulses ($0 \leq t \leq \tau$). Yet, the observation of oscillatory features in the emission signal as a function of the delay between the pulses demonstrates that the exciton wave packets are robust under individual realizations of the environmental fluctuations on a time scale of 250 fs (our maximum delay time), whereas the distributions of damping times $T_d$ in Fig.2D-F testify to the heterogeneity in the environmental fluctuations of the individual complexes that cause the loss of coherence. Each ring of BChl *a* molecules in each complex features its particular realization of the protein environment with its specific fluctuations and thus a specific time scale for the loss of coherence.

For repetitive pulse sequences, an even stricter condition applies for observing modulations in the emitted signal, namely that for a fixed delay time $\tau$ the sum of correlations $\sum_i\langle\delta\omega_i(\tau,0)\cdot\delta\omega_i(\tau,t)\rangle_t \neq 0$ must remain finite over the data acquisition time $t_{av}$ ($0 \leq$ $t \leq t_{av}$)[39]. This is a common consideration in experiments involving repetitive measurements, such as 2DES[56–58]. In practice, however, the magnitude of the dynamic fluctuations $\delta\omega_i(t)$ are usually small compared to the static mean $\left(\omega_0 + \overline{\delta\omega_i}\right)$, so these conditions are typically satisfied, and lead only to a reduction of the maximum achievable amplitude of the modulation (for a more detailed description of the fluctuations on different time scales see section 6 in the supporting information). Once the temporal fluctuations of the mean site energies become too large during the data acquisition time $t_{av}$ the realizations of the energies differ too much from each other and interferences cannot be recorded due to vanishing amplitudes.

For the scenarios A and B, the distributions of the oscillation frequencies, Fig.2A, B, and of the damping times, Fig.2D, E, are equivalent within statistical accuracy. Thus, whether the initial excitation includes (scenario A) or excludes (scenario B) the B800 states does not play a significant role in the relaxation within the B850 manifold. This is not too surprising, because the relaxation between the B800 and B850 manifolds takes place on a time scale of a few picoseconds[59–62], which is much longer than the maximum delay time between the pulses in the current experiments. In contrast, the choice of the lock frequency $\Omega_L$ is very important, since it allows us to probe the B850 relaxation dynamics for different realisations of the wavepackets. The lock frequency determines the spectral components, that drive the system, (see Fig.S2), and thus determines the time-dependent coefficients $c_i(t)$ of the exciton states contributing to the wave packets which ultimately lead to a characteristic modulation of the emitted signal (cf. Fig. 1G and I). As outlined above, this modulation is governed by the time-dependent phase $\phi(t) = -i(\Omega_L - \omega_i)t$ accumulated by each electronic state $|\Psi_i\rangle$ that possesses a finite transition moment to the ground state and is still (partially) occupied when the second pulse arrives - a prerequisite for wave packet interference to be detectable. This contrasts with excitation spectroscopy, that merely probes whether the laser frequency is resonant with the exciton energy.

Since, mainly the lowest-energy B850 states, $k = 0, \pm 1$ carry oscillator strength to the ground state, these states contribute predominantly to the wave packets. The modulation frequency observed in the emission signals is therefore mainly determined by the energy difference between the lock frequency and the energy of these B850 states. This explains the nearly similar mean values of $\overline{f} \approx 700\ cm^{-1}$ of the oscillation frequencies for identical lock frequencies of scenarios A and B, and the significant drop to $\overline{f} \approx 290\ cm^{-1}$ for lowering the lock frequency to an equivalent of 840 nm of scenario C. If the lock frequency is shifted even further to an equivalent of 860 nm, i.e. resonant with the energies of the $k = \pm 1$ states, corresponding to $(\Omega_L - \omega_i(k = \pm 1) \approx 0)$, the contribution from these states to the modulation of the emission is cancelled out. Then contributions to the modulation of the emission that correspond to the difference between the lock frequency and the energies of higher exciton states become visible, although with a small modulation amplitude only, since these higher exciton states carry only little oscillator strengths to the ground state (and their contributions are thus overridden in the other scenarios), see Fig.S7 in section 7 of the supporting information. In any case, the B850 exciton state with the largest (time-dependent) coefficient $c_i(t)$ after relaxation determines the wave packet interference.

Interestingly, for some of the complexes the data reveal temporal variations of the modulation frequencies $f$ on slow time scales (60 s). This is illustrated in more detail in Fig.3A-C for the three scenarios A, B, and C. The vertical green lines refer to the lock frequencies $\Omega_L$ and each horizontal line corresponds to the difference $\Delta\Omega = \Omega_L - f$ between the lock frequency and the modulation frequency. Each colour (black, red, blue) corresponds to the results from a particular individual complex. For ease of comparison, we overlay the spectrum of the excitation laser, the ensemble absorption spectrum featuring the B800/B850 bands, and the spectral positions of the energies of the B850 exciton states that were obtained from single complexes at 1.4 K by some of us in previous work, see also Fig.S8 in section 8 of the supporting information[63]. Notably, for all complexes $\Delta\Omega$ fluctuates around the spectral positions of the lowest B850 exciton states, $k = 0$ and $k = \pm 1$, i.e. those states that carry most of the oscillator strength to the ground state.

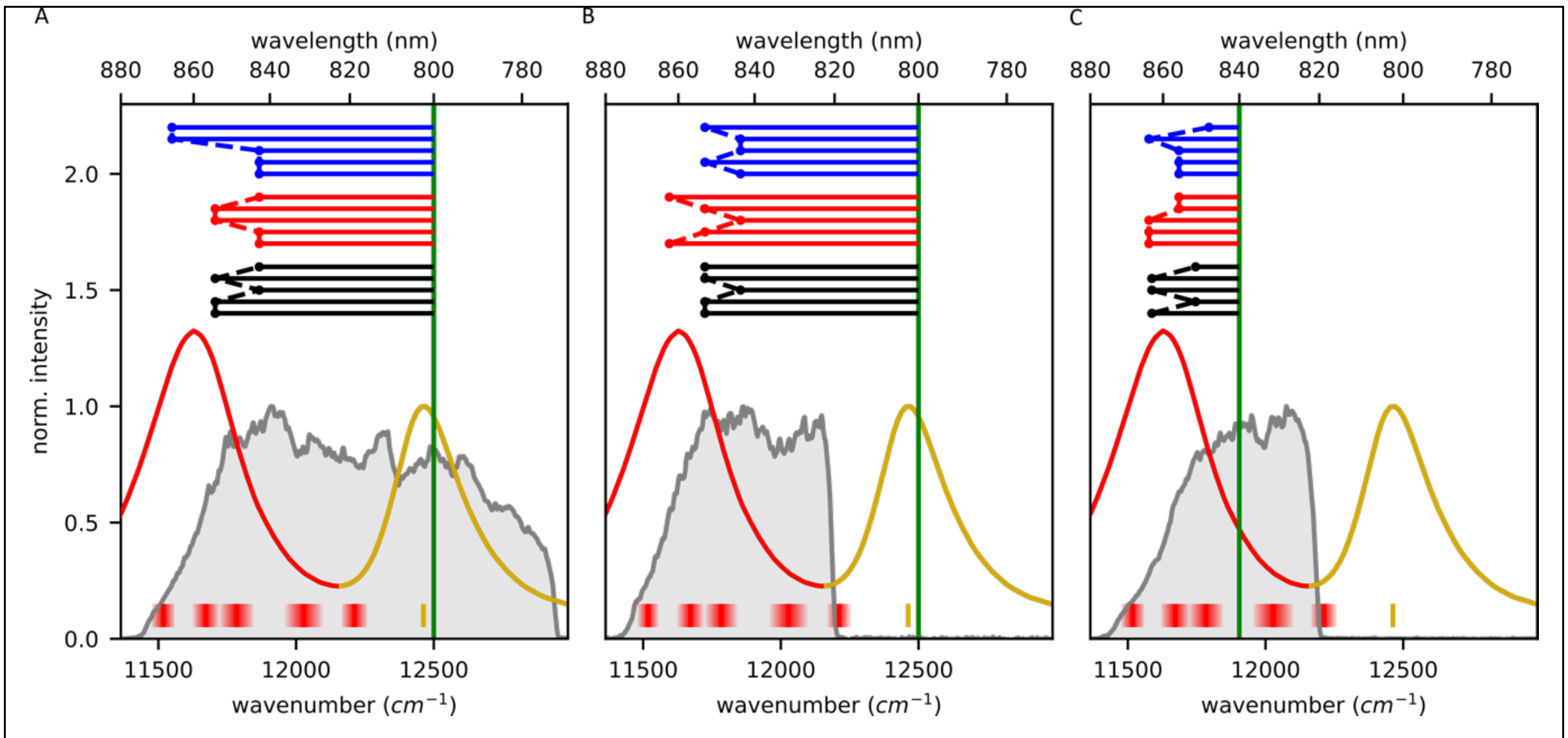

Fig.3, Bottom: The red bands refer to the spread of the energies of the lowest B850 exciton states, as obtained from single LH2 complexes from *Rps. acidophila* at 1.4 K, see also Fig.S8[63]. For ease of comparison the corresponding ensemble absorption spectrum and the laser spectrum have been overlaid. Top: Differences between the lock frequency (green vertical line) and the modulation frequencies, $\Delta\Omega = \Omega_L - f$, for individual scans (dots) of individual LH2 complexes from *Rps. acidophila*. Each colour corresponds to the results that are associated with one particular complex. The duration of a single scan was 60 s. Parts A) - C) of the figure correspond to scenarios A - C, respectively.

Since our approach creates wave packets comprising only excited states carrying oscillator strength with the ground state, the agreement of spectral positions of the excited states with low-temperature data demonstrate that their spectral positions are largely independent of temperature. More importantly, however, these data show that the population of states contributing to wave packet interference change on time scales of tens of seconds. This implies that relaxation pathways to the $k = 0, \pm 1$ states change as well on those timescales.

Accordingly, the gap in the energies between the participating exciton states has to be bridged by bath fluctuations. It is commonly accepted that energy transfer in photosynthetic pigment-protein complexes is most efficient in the intermediate regime where the coupling $V$ between the chromophores is of the same order as the reorganization energy $\lambda$[12,64,65]. The reorganization energy can be expressed as $\lambda = S\,h\nu_{vib}$, where $S$ denotes the Huang Rhys (HR) factor, $h$ Planck's constant, and $\nu_{vib}$ the vibrational frequency of the mode that couples to the electronic states. For the transitions from emitting B850 states the HR factor has been measured by differential fluorescence line narrowing and varied between 1 and 2 as a function of the excitation wavelength[66]. Theoretical modelling that included exciton self-trapping predicted variations between 0.4 and 2.8[67], which was later confirmed by single-molecule techniques[68,69]. This HR factor reflects electron-phonon coupling in the ground state but is reasonable to assume that the HR factor in the excited states is of similar magnitude. Since for LH2 from purple bacteria $V \approx 200 - 400$ $cm^{-1}$ holds for the intermolecular coupling within the B850 ring[37], the intermediate coupling regime implies that mainly low-frequency vibrations with energies around 100 - 300 $cm^{-1}$ are coupled to the electronic states. This is in line with the lack of vibrational bands of higher frequencies in the emission spectra from single LH2 complexes[68,70]. Moreover, it was uncovered that the HR factors for an individual complex can fluctuate between the aforementioned extremes on a time scale of several seconds[69]. Therefore, we suggest that these changes in relaxation pathways reflect structural fluctuations in the vicinity of the binding pockets of the BChl *a* molecules, which is consistent with previous observations of structural (conformational) fluctuations of single LH2 complexes on similar time scales[18,20–22,38,71,72]. These fluctuations impact on the site energies and/or the mutual electronic coupling strength between the BChl *a* molecules, with corresponding consequences for the excited state energy landscape.

For several light-harvesting antenna complexes, including LH2, it has been found that the energy gaps between the exciton states match with energies of specific intra-molecular vibrational states[27,73–76], and it has been hypothesized that this is not a coincidence, but that this enables the electronic system (excitons) to harvest nuclear fluctuations for efficient relaxation between electronic states[27,57]. At the same time, this coupling between electronic states and (intra-molecular) vibrations allows exciton states to borrow coherence from vibrations, which was proposed to give rise to the long-lived oscillatory signals observed in 2DES experiments (although the origin of those long-lived oscillations is very likely related to vibrational/vibronic coherences in the electronic ground state that are not relevant for the function of light-harvesting complexes[25,77]). Recently, in the case of LH2 (from *Rps. acidophila*) the idea that efficient energy relaxation in the B850 manifold of exciton states relies on specific fine-tuned electron-vibrational resonance conditions has been challenged by 2DES studies[78]. In agreement with the results reported in[78] the findings of the present study indicate that Nature has instead preserved an electronic energy landscape for light harvesting that is sufficiently robust to structural disorder by relying on the many available low-frequency intra- and intermolecular vibrations (with energies below 300 $cm^{-1}$), to ensure efficient relaxation to the emissive B850 state via specific $k$ states for each complex at any given moment.

## Conclusion

The current study has allowed us to combine complementing information from time-resolved spectroscopy and results from earlier linear spectroscopy, linking fluctuations of the exciton relaxation pathways to the energies of the involved exciton states. This provides deeper insights into the dynamics of the energy relaxation within the B850 exciton states of individual LH2 complexes.

## Acknowledgment

S.W., P.R. and J.K. thankfully acknowledge financial support by the Deutsche Forschungsgemeinschaft (DFG, KO1359/31-1) and the State of Bavaria within the initiatives "Solar Technologies go Hybrid" as well as the Elite Network "Biological Physics". We thank Ralf Kunz who determined the energies of the B850 exciton states.

## Author information

ORCID:

| | |
|---|---|
| R.J. Cogdell | 0000-0003-2119-6607 |
| A. Gardiner | 0000-0001-6161-2914 |
| R. Hildner | 0000-0002-7282-3730 |
| J. Köhler: | 0000-0002-4214-4008 |

**Supporting Information for**

# Fluorescence-detected Wavepacket Interferometry reveals time-varying Exciton Relaxation Pathways in single Light-Harvesting Complexes

Stephan Wiesneth[1], Paul Recknagel[1], Alastair T. Gardiner[3]
Richard Cogdell[2], Richard Hildner[4], Jürgen Köhler[1,5,6]

[1]*Spectroscopy of soft Matter, University of Bayreuth, 95440 Bayreuth, Germany*
[2]*School of Molecular Biosciences, Glasgow University, Glasgow G12 8QQ, UK.*
[3] *Institute of Microbiology of the Czech Academy of Sciences, 379 01 Třeboň, Czech Republic*
[4]*Zernike Institute for Advanced Materials, University of Groningen, Nijenborgh 3, 9747 AG Groningen, The Netherlands.*
[5]*Bavarian Polymer Institute, University of Bayreuth, 95440 Bayreuth, Germany*
[6]*Bayreuther Institut für Makromolekülforschung (BIMF), 95440 Bayreuth, Germany*

**Table of Content**

## 1. Structure of LH2

The atomistic structures of the peripheral light-harvesting complexes (LH2) from purple bacteria depend on the bacterial species and on the growth conditions[1–8]. An overarching structural feature is a unit of two transmembrane $\alpha$ and $\beta$ apoproteins that accommodate a small number of molecules of BChl *a* and carotenoid. These $\alpha\beta$ subunits oligomerize, and form a complete ring for the LH2 complexes Fig.S1A. The BChl *a* molecules within LH2 can be grouped into two pigment pools that are arranged in two concentric rings that are shifted with respect to each other along a common symmetry axis. Within the two BChl *a* pools the intermolecular interaction strengths differ significantly resulting in one ring of weakly interacting BChl *a* molecules (one BChl *a* molecule per basic protein heterodimer), and one ring of closely interacting BChl *a* molecules (two BChl *a* molecules per basic protein heterodimer), which for the species *Rps. acidophila*, give rise to two absorption bands at 800 nm and 850 nm, respectively, Fig.S1B.

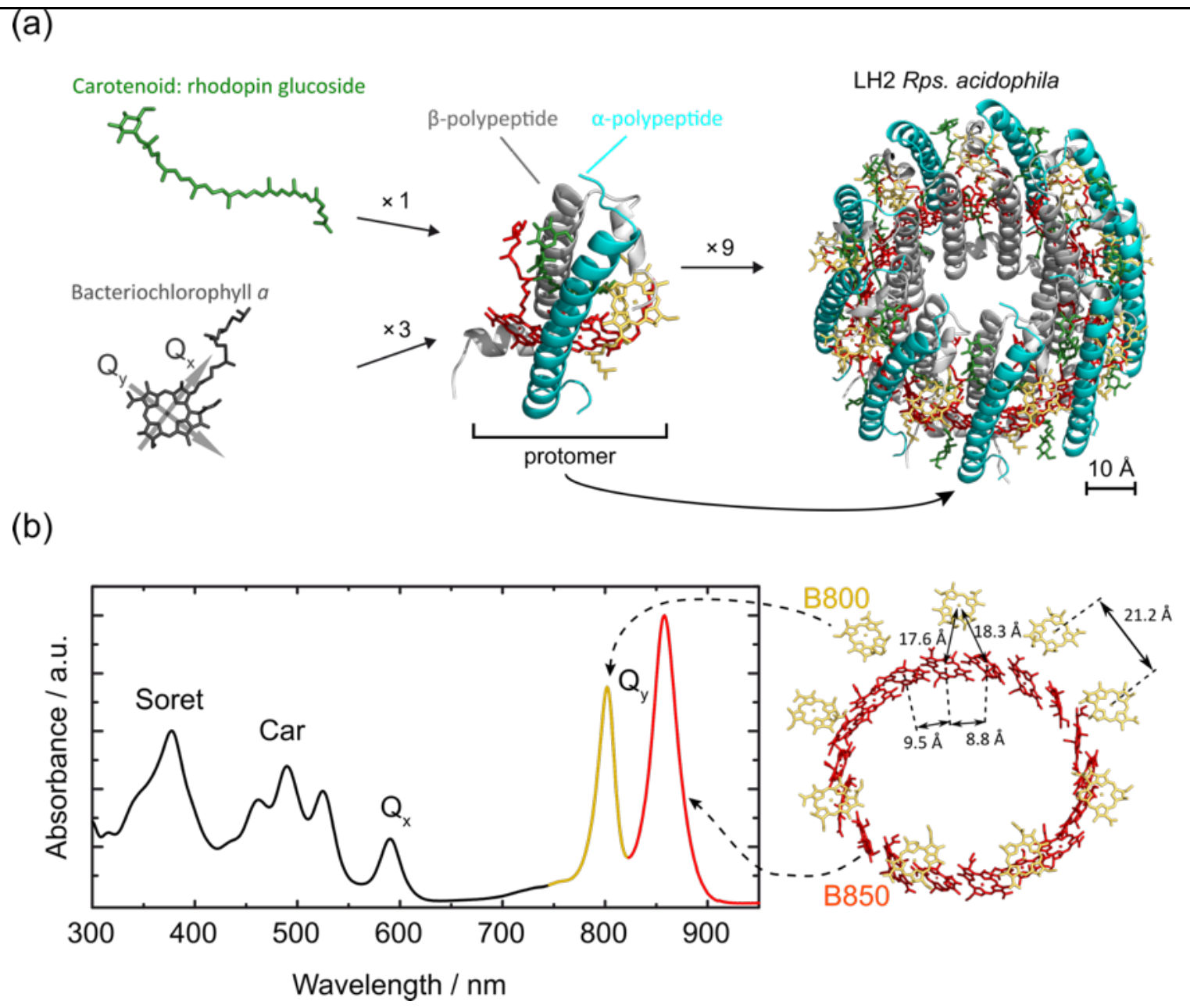

Fig.S1: General construction principle for the peripheral light-harvesting complexes (LH2) from purple bacteria illustrated on the example of the B800-B850 LH2 complex from the species *Rps. acidophila*. Adapted from[9]. A) One molecule of carotenoid (here rhodopin glucoside) and three molecules of BChl *a* are kept together by two apoproteins ($\alpha$, $\beta$, polypeptide, white and blue, respectively) to form a protomer. The different colours of the BChl *a* molecules indicate the association with the B800 (yellow) and B850 (red) pigment pools. Nine protomers oligomerize and form the LH2 complex of *Rps. acidophila*. B) Absorption spectrum from an ensemble of B800-B850 LH2 complexes. The B800 and B850 bands are colour coded in yellow and red, respectively. The arrangement of the BChl *a* molecules in the B800 ring (yellow), and in the B850 ring (red) is shown next to the spectrum together with the centre-to-centre separations of the BChl *a* molecules in Å[10].

## 2. Amplitude and phase shaping

Phase-locked ultrashort pulses have been generated in a spatial light modulator applying an amplitude modulation mask and a phase mask. Examples for pulse pairs and the corresponding masks are shown in Fig.S2.

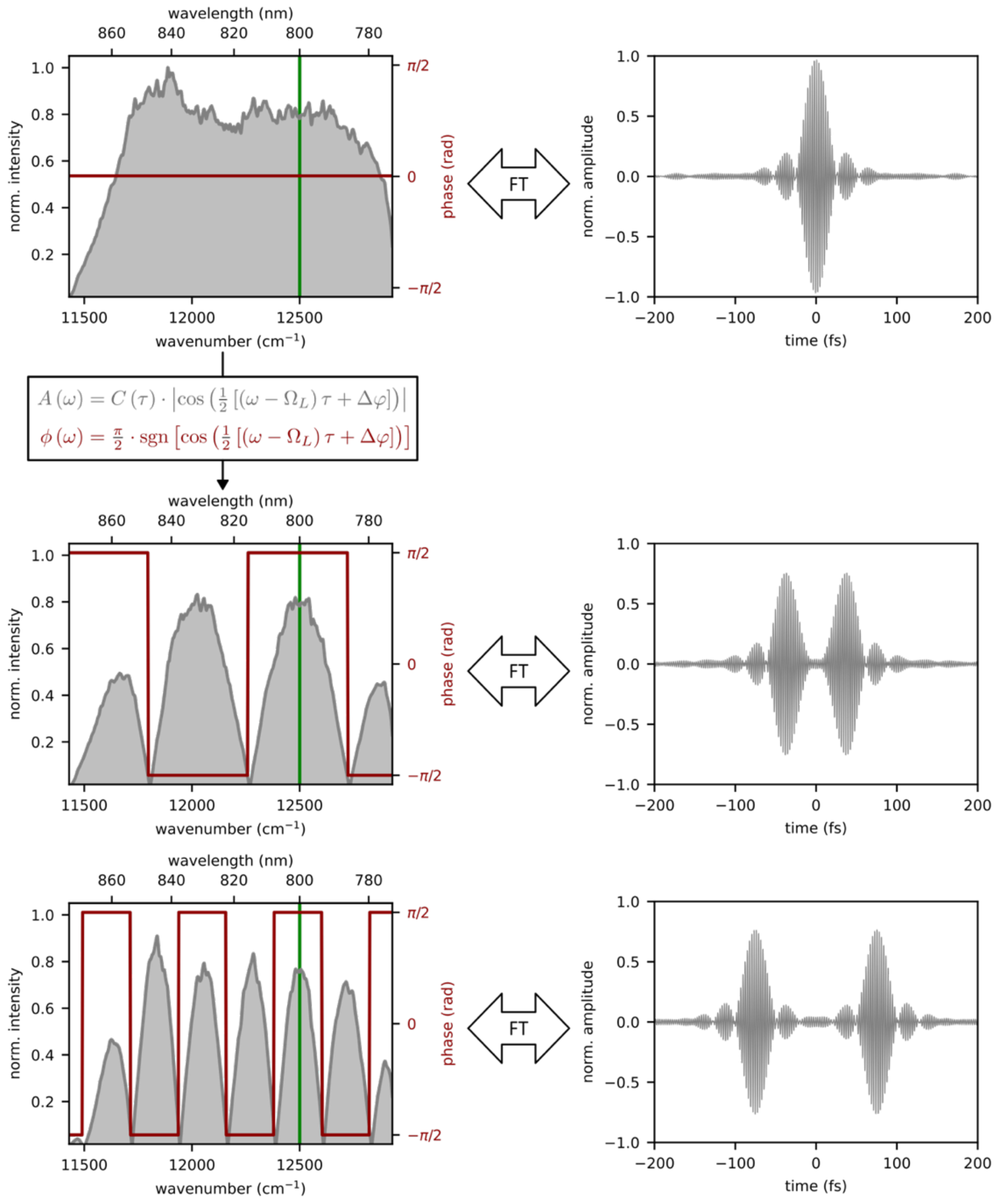

Fig.S2: Examples for the generation of phase-locked pulse pairs separated by $\tau = 72$ fs and $\tau = 150$ fs, respectively, using an amplitude modulation mask and a phase mask. Top left: Initial laser spectrum (grey area). Centre and bottom left: The laser spectrum is subjected to an amplitude modulation mask $A(\omega) = C(\tau)|cos(0.5[(\omega - \Omega_L)\tau + \Delta\varphi])|$ and a phase mask $\phi(\omega) = \frac{\pi}{2} sgn[cos(0.5[(\omega - \Omega_L)\tau + \Delta\varphi])]$ that alternates the sign of the phases (green) for the spectral components. The blue lines indicate the lock frequency. Right: The resulting Fourier transforms (FT) in the time domain.

### 3. Single-step photobleaching

Integrated emission intensity from LH2 as a function of time. The one-step bleaching process is considered as evidence for dealing with an individual complex.

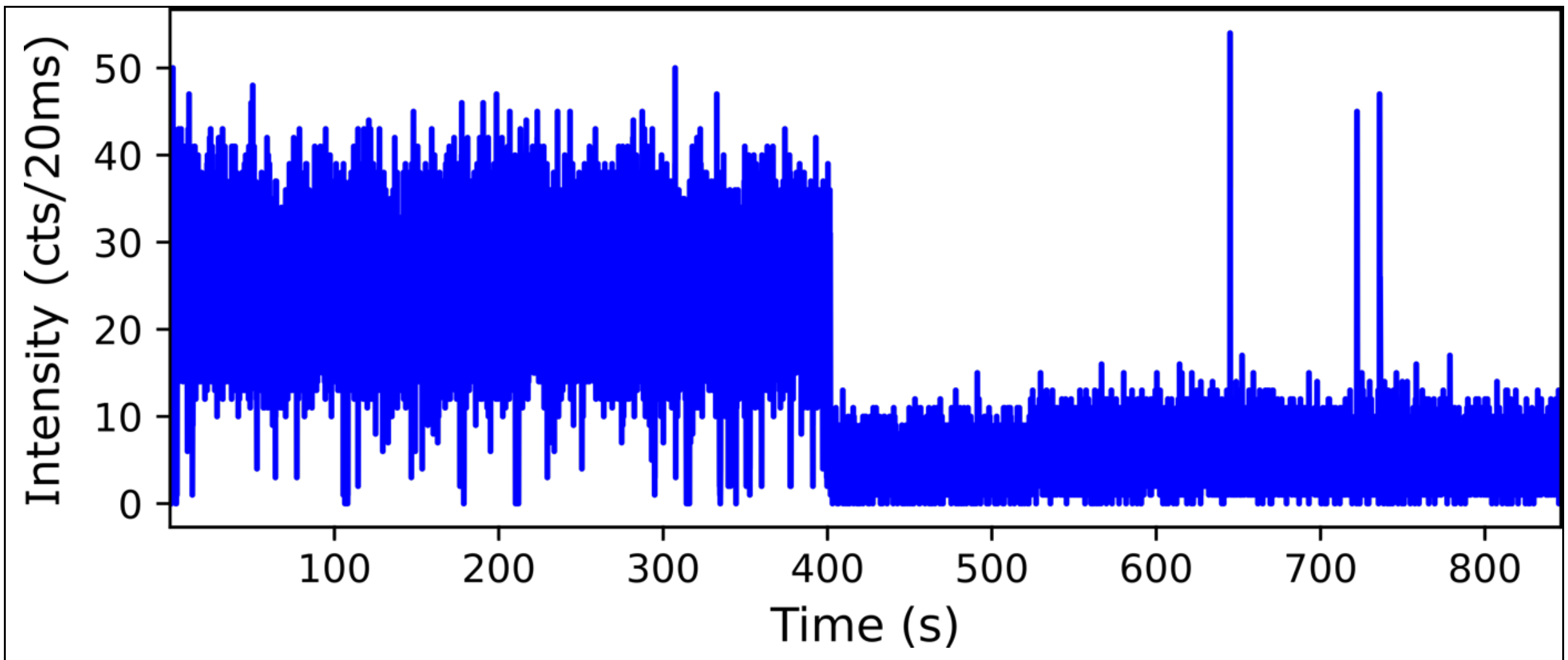

Fig.S3: Emission intensity recorded from a single LH2 complex. The signal is given in counts per 20 ms. The bin time was 20 ms and the excitation intensity was 300W/cm$^2$.

## 4. Example of a Fourier transform

Illustration of the Fourier transforms from a sequence of time traces from an individual LH2 complex.

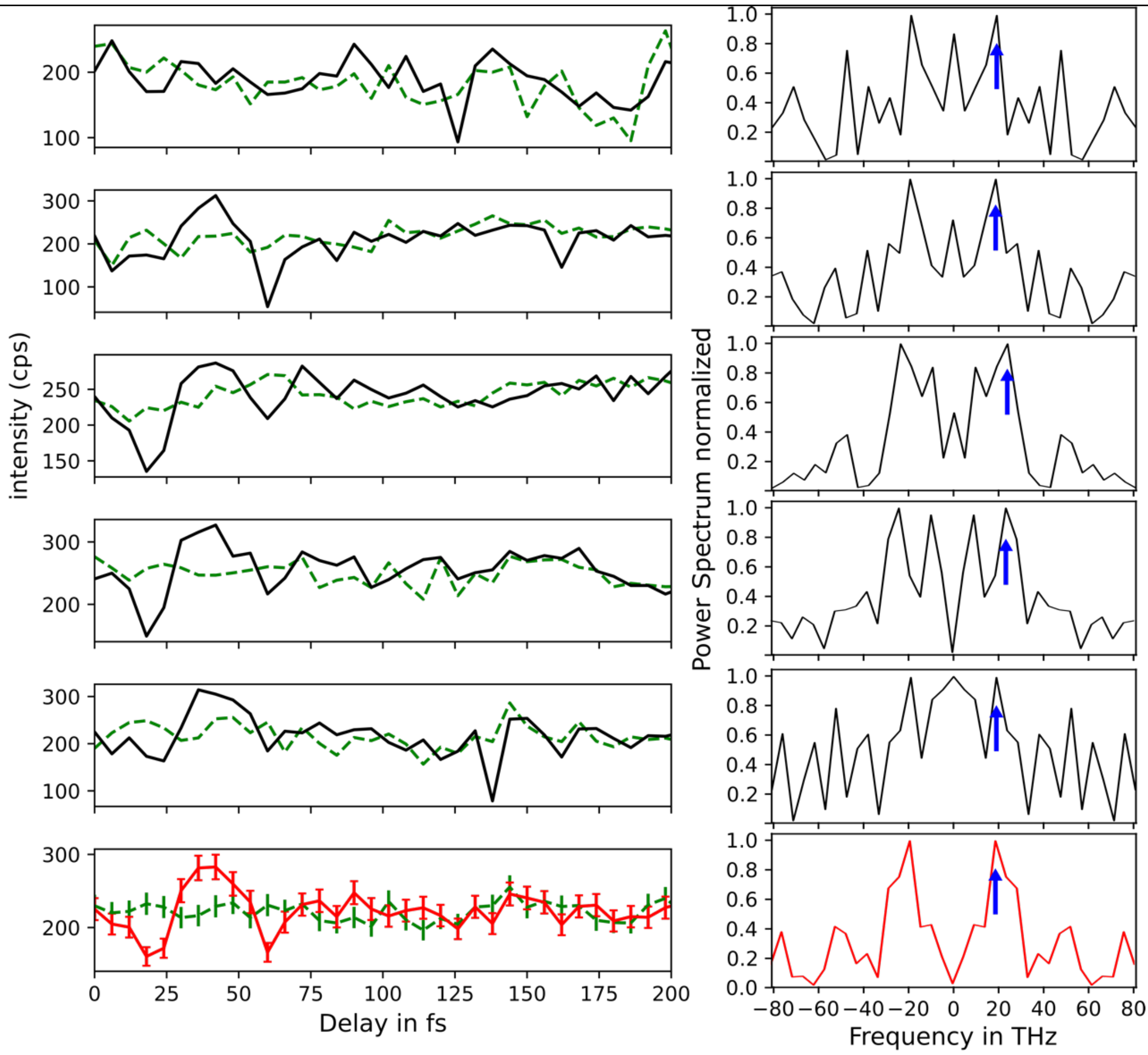

Fig.S4: Left: Response from a single LH2 complex upon excitation with two phase-locked pulses as a function of the pulse delay $\tau$ for the excitation conditions corresponding to scenario A (lock frequency corresponding to 800 nm, excitation intensity 200 W/cm$^2$, bin time 1s). The black curves are individual scans with a phase difference of $\Delta\varphi = 0$ between the pulses, and the green lines correspond to the reference signal obtained from a single transform-limited pulse. The red trace at the bottom corresponds to the average of the black traces. The error bars take the shot noise and fluctuations of the laser intensity into account. Right: Fourier transforms of the individual transients (black) and of the averaged transient (red). The blue arrows indicate the extracted Fourier frequencies.

## 5. Illustration of wavepaket interference

Schematic representation of the processes in a single LH2 complex upon excitation with two phase-locked laser pulses[11]. For simplicity only three essential states are considered, namely a B800 state ($|\Psi_{800}\rangle$), a higher exciton state in the B850 band ($|\Psi_{850}^{h}\rangle$), and an emitting state in the B850 band ($|\Psi_{850}^{e}\rangle$).

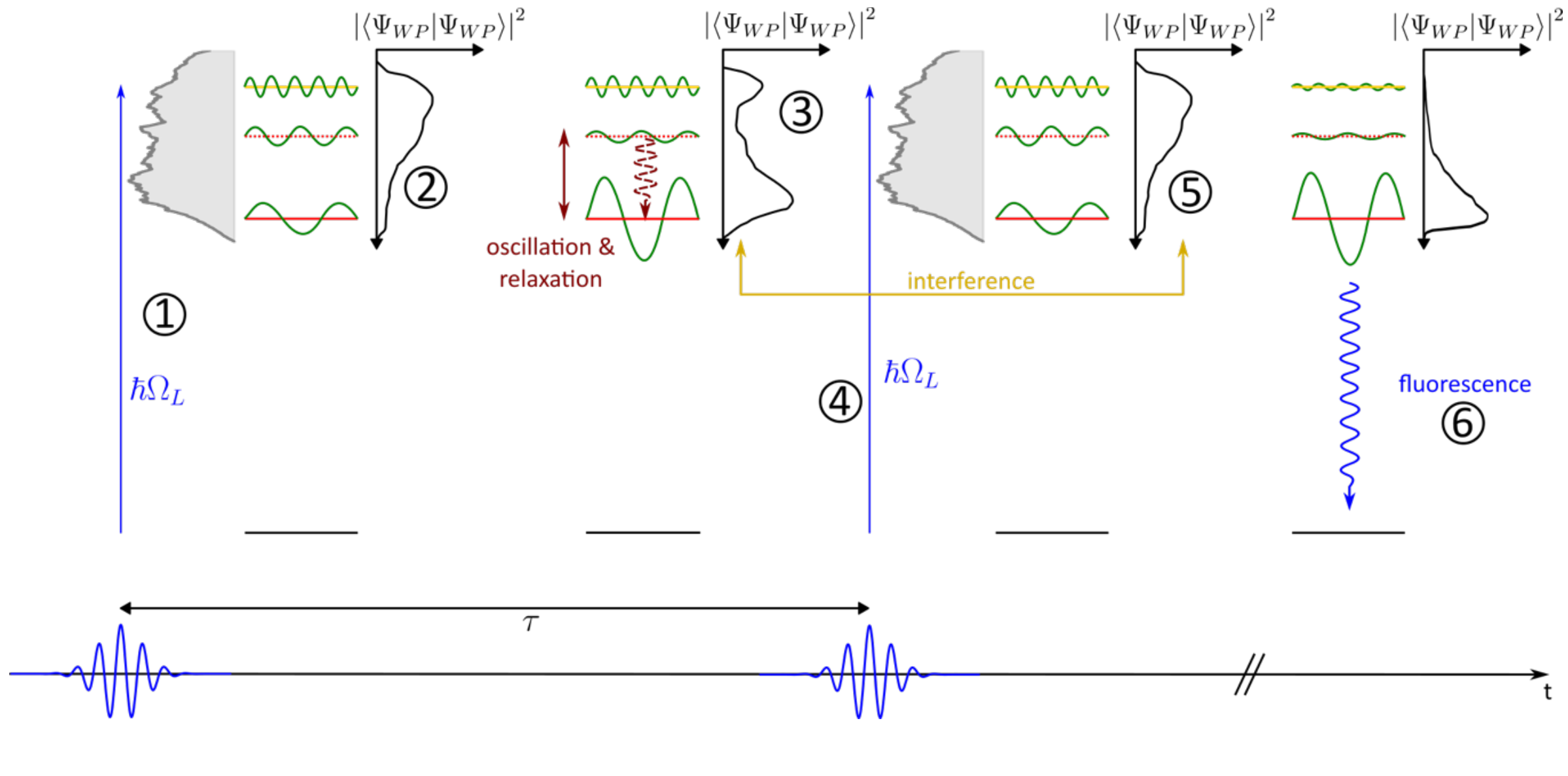

Fig.S5: Simplified energy level diagram that represents the electronically excited states of a single LH2 complex. It distinguishes between $|\Psi_{800}\rangle$ (yellow line), $|\Psi_{850}^{h}\rangle$(dashed red line) and $|\Psi_{850}^{e}\rangle$ (full red line). The bottom line represents a time axis with the two pulses separated by a delay time τ. ① For scenario A the excitation with the full laser spectrum creates an exciton wavepacket $|\Psi_{WP}(t)\rangle = c_{800}(t)|\Psi_{800}\rangle + c_{850}^{h}(t)|\Psi_{850}^{h}\rangle + c_{850}^{e}(t)|\Psi_{850}^{e}\rangle$. The laser spectrum is indicated in grey on the left hand side of the scheme. The amplitudes of the individual energy states are indicated by the green oscillating lines. ② Refers to the probability $|\langle\Psi_{WP}|\Psi_{WP}\rangle|^2$ immediately after absorption of the first pulse. ③ During the delay time $\tau$ the composition of the wavepacket evolves due to oscillation and relaxation. ④ The second pulse launches a similar wavepacket as the first pulse. The two laser pulses are phase locked at the frequency $\Omega_L$ that is indicated by the blue vertical arrows. ⑤ The two wavepackets interfere and the time-dependent population of the emitting states determines the fluorescence intensity ⑥.

**6. Timescales and averaging**

The energies of the states $|\Psi_i(t)\rangle$ can be expressed as $\hbar\omega_i = \hbar\left(\omega_0 + \overline{\delta\omega_i} + \delta\omega_i(t)\right)$, which takes temporal fluctuations with amplitudes $\delta\hbar\omega_i(t)$ around a static average $\hbar\left(\omega_0 + \overline{\delta\omega_i}\right)$ into account. This yields for the phases $\phi(t) = -i\left(\Omega_L - \left(\omega_0 + \overline{\delta\omega_i} + \zeta + \delta\omega_i(t)\right)\right)t$ and $\phi(t+\tau) = -i\left(\left(\Omega_L - \left(\omega_0 + \overline{\delta\omega_i} + \delta\omega_i(t+\tau)\right)\right)(t+\tau) + \varphi\right)$, and interferences can be observed as long as the sum of the correlations of the fluctuations $\sum_i\langle\delta\omega_i(0)\cdot\delta\omega_i(\tau)\rangle_t \neq 0$ does not vanish. The index *t* refers to a temporal averaging, and the result of the autocorrelation depends on the time duration that is used for the averaging procedure.

The temporal average of the fluctuations over the pulse length $\tau_{pulse}$ is given as $\langle\delta\omega_i(t)\rangle_{\tau_{pulse}} = \frac{1}{\tau_{pulse}}\int_t^{t+\tau_{pulse}}\delta\omega_i(t')dt'$. Under the assumption $\delta\omega_i(t) \approx \delta\omega_i(t+\tau_{pulse}) \approx \delta\omega_i\left(\mathrm{t} + \frac{\tau_{pulse}}{2}\right)$, we can approximate $\langle\delta\omega_i(t)\rangle_{\tau_{pulse}} \approx \delta\omega_i\left(t + \frac{\tau_{pulse}}{2}\right)$, and define $\overline{\delta\omega_i}(\tau,t)$ as the average of the fluctuations over the pulse separation $\tau$ at (the macroscopic) time $t$, i.e., $\overline{\delta\omega_i}(\tau,t) := \left\langle\delta\omega_i\left(t + \frac{\tau_{pulse}}{2}\right)\right\rangle_\tau = \frac{1}{\tau}\int_t^{t+\tau}\delta\omega_i\left(t' + \frac{\tau_{pulse}}{2}\right)dt'$, see Fig.S6A. Then a prerequisite for observing interferences between the wavepackets is that the realizations of $\delta\omega_i\left(t + \frac{\tau_{pulse}}{2}\right)$ do not differ too much from each other within the lag time $\tau$ between the pulses. In other words, as long as the sum of the autocorrelations fulfills $\sum_i\langle\delta\omega_i(0)\cdot\delta\omega_i(t)\rangle_t \neq 0$, where the brackets refer to the average over time *t* that runs from $0 \leq t \leq \tau$.

However, since a single two-pulse experiment will not give a detectable signal the pulse sequences are applied repetitively, and for a fixed delay time $\tau$ the signal obtained corresponds to $\left\langle\overline{\delta\omega_i}(\tau,t)\right\rangle_{t_{av}} = \frac{1}{t_{av}}\int_t^{t+t_{av}}\overline{\delta\omega_i}(\tau,t')dt'$, where $t_{av}$ denotes the data acquisition time, Fig.S6B. Then the next prerequisite for observing interferences between the wavepackets that should be fulfilled is that the averages $\overline{\delta\omega_i}(\tau,t)$ should not differ to much from each other during the data acquisition time $t_{av}$ [11]. Or in other words as long as $\left(\omega_0 + \overline{\delta\omega_i}\right) \gg \delta\omega_i(t)$ holds, which is usually fulfilled for the pigment-protein complexes. Finally, the interference trajectories correspond then to $\langle\overline{\delta\omega_i}(\tau,t)\rangle_{t_{av}}$ displayed as a function of the delay time $\tau$ between the pulses, Fig.S6C.

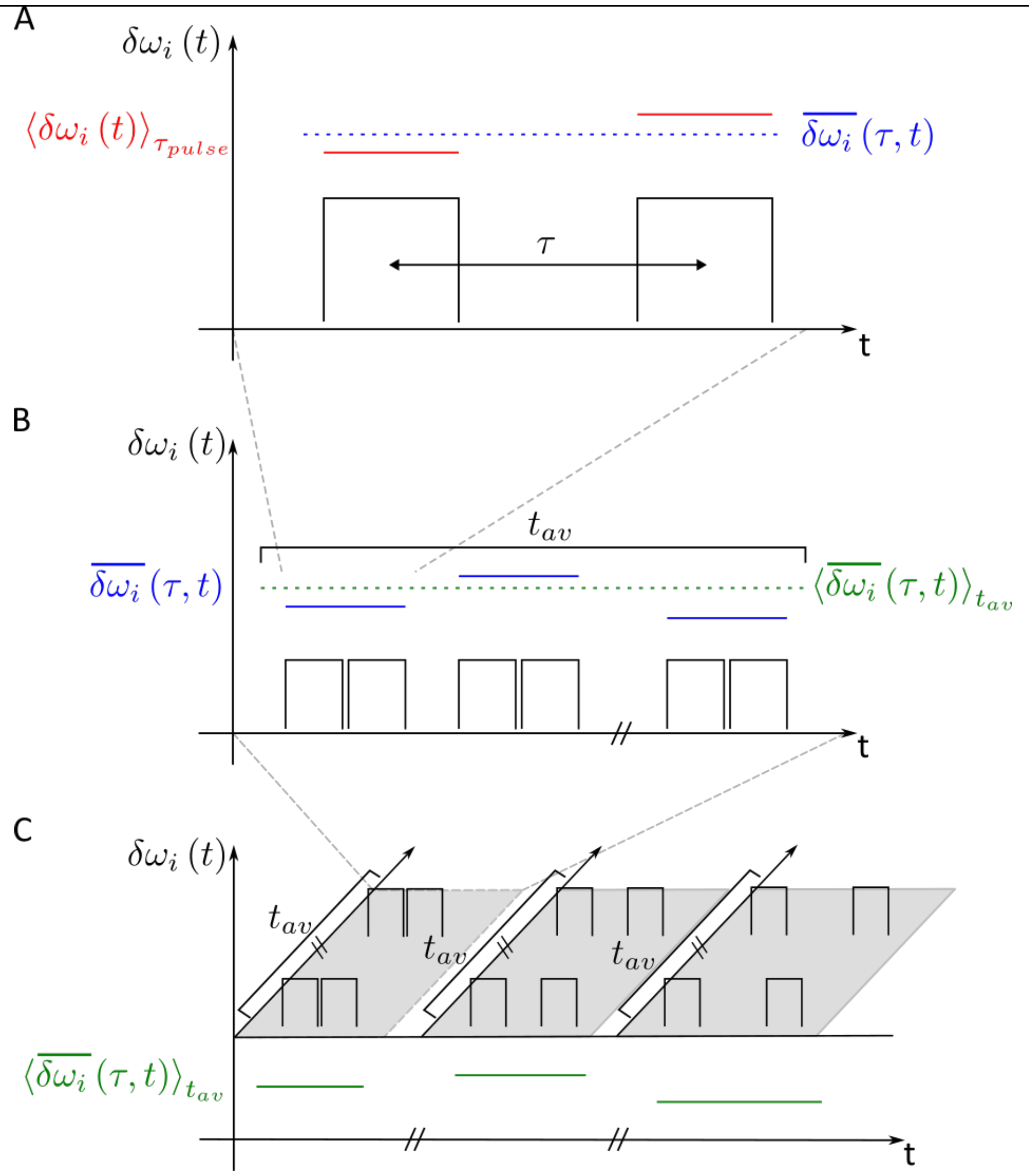

Fig.S6: Schematic illustration of the different time scales that are considered for averaging the fluctuations of the energies of the exciton states. A) Average of the fluctuations $\langle\delta\omega_i(t)\rangle_{\tau_{pulse}}$ during the ultrashort pulse duration (red lines), together with the average $\overline{\delta\omega_i}(\tau,t)$ (blue dotted line) of these fluctuations averaged over a single pulse pair applied at time $t$. B) Average $\langle\overline{\delta\omega_i}(\tau,t)\rangle_{t_{av}}$ (green dotted line) of the fluctuations $\overline{\delta\omega_i}(\tau,t)$ (blue lines) over the time $t_{av}$ that is required for registering one data point for a fixed pulse delay $\tau$. C) Variation of these averages ($\langle\overline{\delta\omega_i}(\tau,t)\rangle_{t_{av}}$, (green lines) for different pulse delays.

## 7. B850-only excitation with a lock frequency equivalent to 860 nm

Comparison of the response from single LH2 complexes using a lock frequency $\Omega_L$ equivalent to 860 nm, i.e., in the spectral range of the lowest B850 exciton states.

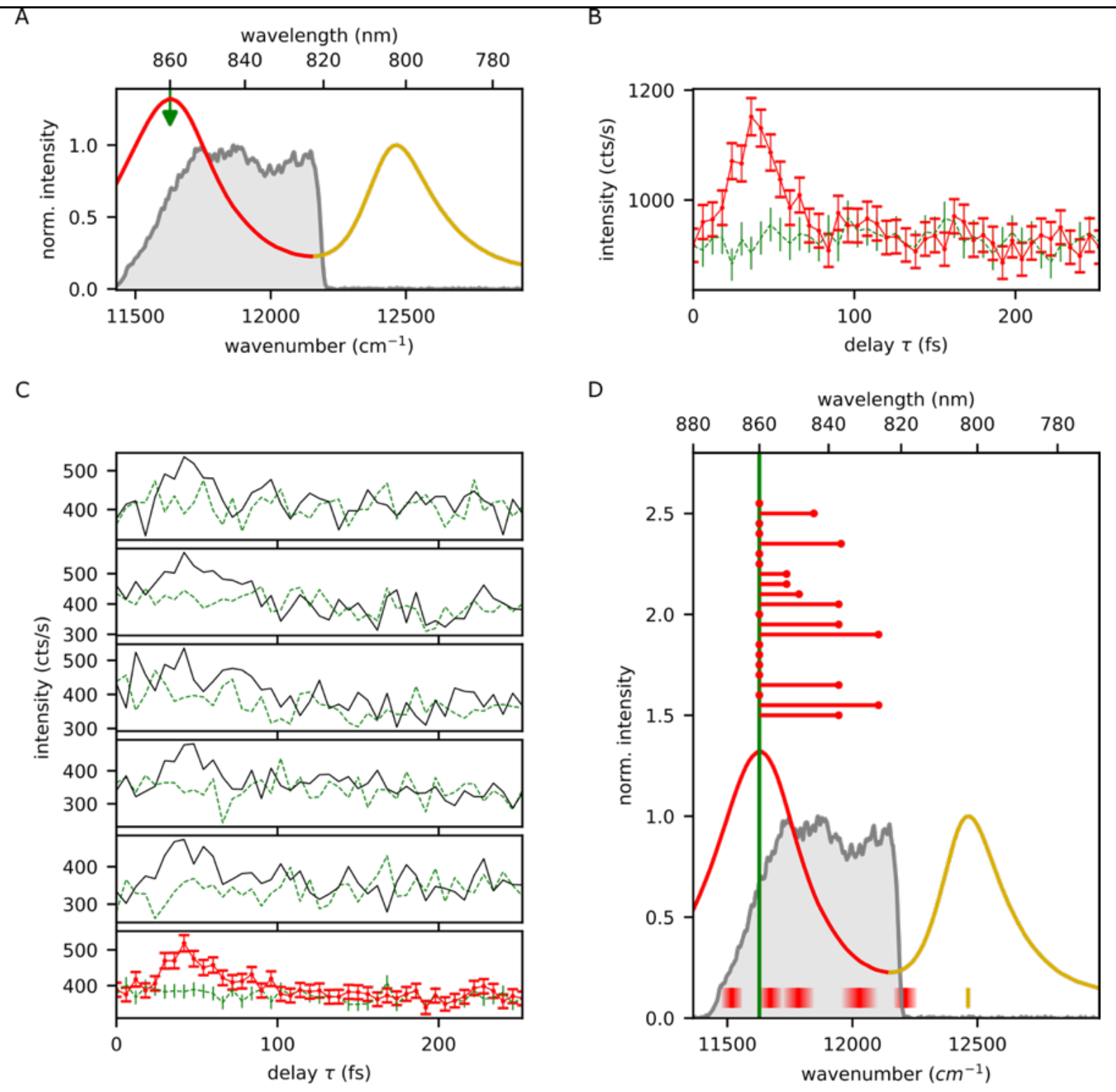

Fig.S7: A) The grey shaded area represents the spectral profile of the laser pulse that has been used for the experiments underneath. The lock frequency for which the phases between the two pulses are locked corresponds to an equivalent of 860 nm and is indicated by the green arrow. For the ease of comparison an ensemble absorption spectrum of LH2 from *Rps. acidophila* has been overlaid, where the yellow (red) part refers to the B800 (B850) absorption band. B) Response from a macroscopic ensemble of LH2 complexes upon excitation with two phase-locked pulses as a function of the pulse separation τ for the excitation conditions shown in part A). C) Response from a single LH2 complex upon excitation with two phase-locked pulses as a function of the pulse delay τ for the excitation conditions shown in part A). The black curves are individual scans, and the red trace at the bottom corresponds to the average of the black traces shown in the same panel. The green lines correspond to the reference signal obtained from a single transform-limited pulse. The error bars take the shot noise and fluctuations of the laser intensity into account. For all experiments the excitation intensity was $200\frac{W}{cm^2}$ during data acquisition. The signal is given in counts per second (cps). The acquisition time for a single scan was 60 s. D) Bottom: The red bands refer to the spread of the energies of the lowest B850 exciton states, as obtained from single LH2 complexes from *Rps. acidophila* at 1.4 K, see also Fig.S8 [9]. For ease of comparison the corresponding ensemble absorption spectrum and the laser spectrum has been overlaid. Top: Difference between the lock frequency (green vertical line) and the mean of the modulation frequency (red line), $\Omega_L - \bar{f}$, from individual LH2 complexes.

## 8. Spectral positions of the lowest B850 states

Illustration of the energetic disorder in the lowest B850 states of LH2 from *Rps. acidophila*. The spectral positions of the exciton states have been determined with single molecule techniques at 1.4 K by some of us[9]. The distributions of the spectral positions are shown in Fig.S8A, and the corresponding means and standard deviations are given in Table.S1. The attribution of particular data points to specific individual complexes is visualized in Fig.S8B. This reveals that the energies of the individual exciton states can vary independently from each other.

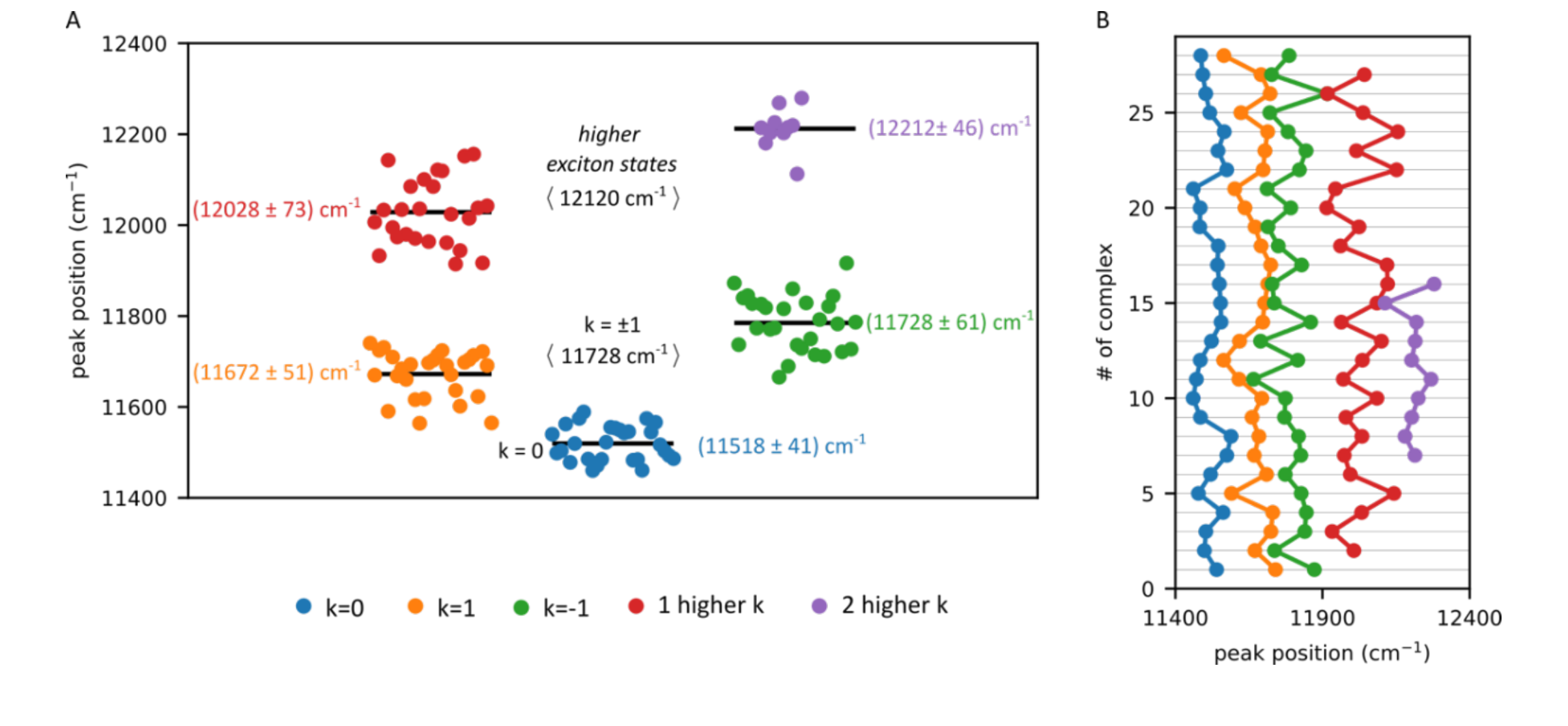

Fig.S8: A) Spread of the energies of the lowest B850 states for single LH2 complexes from *Rps. acidophila*. The means and standard deviations of the distributions are given next to the grey bars whose position indicates the mean value. B) Spectral positions of the lowest B850 states of LH2 for the same individual complexes[9].

In[12] the spectral positions of the B850 states have been modelled taking their mixing with charge-transfer states into account. Comparing the measured and calculated energy differences between adjacent exciton states yields that the modelling can reproduce the qualitative trend of the energy differences, see Table.S1.

Table S1: Comparison of the average spectral positions and spectral separations of the lowest B850 states from single molecule spectroscopy[9] and from theoretical modelling that includes charge-transfer states[12]. The colour code is in accordance with Fig.S8. The statistics of the calculated data have been provided as a courtesy from the authors of[12].

| | experimental | | modelling | |
|---|---|---|---|---|
| | spectral position (mean ± sdev) ($cm^{-1}$) | $\Delta E$ ($cm^{-1}$) | spectral position (mean ± sdev) ($cm^{-1}$) | $\Delta E$ ($cm^{-1}$) |
| k = 0 | 11518 ± 41 | 0 | 11973 ± 85 | 0 |
| k = ± 1 | 11672 ± 51 | 154 | 12082 ± 64 | 109 |
| | 11784 ± 61 | 112 | 12157 ± 52 | 75 |
| k = ± 2 | 12028 ± 73 | 244 | 12282 ± 50 | 125 |
| | 12212 ± 46 | 184 | 12348 ± 44 | 66 |